\documentclass[aps,prl,twocolumn,superscriptaddress,preprintnumbers,showpacs]{revtex4}
\usepackage{graphicx,color}
\begin{document}

\title
{ Remarks on the 2nd order Seiberg-Witten maps}

\author{Josip Trampeti\'{c}}
\affiliation{Theoretical Physics Division, Rudjer Bo\v skovi\' c Institute, 
Zagreb, Croatia}
\author{Michael\ Wohlgenannt}
\affiliation{Erwin Schr\"odinger International Institute for Mathematical Physics,
  Boltzmanngasse 9, 1090 Wien, Austria}

\date{\today}

\begin{abstract}
In this brief report, we discuss the Seiberg-Witten maps up to the second order
in the noncommutative parameter $\theta$. They add to the recently 
published solutions in \cite{Alboteanu:2007bp}.
Expressions for the vector, fermion and Higgs fields are given explicitly.
\end{abstract}

\maketitle

The main purpose of this brief report is to complete 
the second order Seiberg-Witten (SW) maps 
constructed in  \cite{Alboteanu:2007bp,Moller:2004qq}. 
We consider canonically deformed space-time. 
The commutator of
coordinates is given by the constant antisymmetric matrix $\theta^{\mu\nu}$,
\begin{equation}
[x^\mu \stackrel{\star}{,} x^\nu] \equiv 
x^\mu \star x^\nu - x^\nu \star x^\mu = i \theta^{\mu\nu}\,,
\label{sw1}
\end{equation}
where we have used the Weyl-Moyal star product
\begin{equation}
(f\star g)(x) = \exp \left\{ \frac{i}2 \theta^{\mu\nu} 
\frac{\partial}{\partial y^\mu} \frac{\partial}{\partial x^\nu} \right\}
f(y)g(x)\bigg|_{y\to x}\,.
\label{sw2}
\end{equation}

A prescription for constructing arbitrary gauge theories on 
a NC space-time was presented in \cite{Seiberg:1999vs}.
Seiberg-Witten maps \cite{Seiberg:1999vs,Madore:2000en} relate
noncommutative (NC) gauge fields and ordinary fields in commutative
theory  via a power series expansion in $\theta$. 
In simplest possible approach to the construction of NC gauge field theories 
all field products are replaced by $\star$-products. This approach, however, 
fails for general gauge theories. For example for SU($N$) gauge theories,
the $\star$-commutator of two infinitesimal gauge transformations 
does not close in the SU($N$) Lie algebra.
This is the reason one has to go to the enveloping algebra \cite{Jurco:2000ja} 
of the Lie algebra of a given group. Higher-order SW
terms are now expressed in terms of the zeroth-order (commutative) fields, 
and as a consequence we do
have the same number of degrees of freedom as in the commutative case. 

The SW maps are not unique. The free parameters are chosen
such that the non-commutative gauge fields are hermitian and the
action is real. Still, there is some remaining freedom including the
freedom of classical field redefinition and noncommutative gauge transformation.

As already remarked in \cite{Alboteanu:2007bp}, the second order solution 
for the gauge field - and therefore also
for the field strength - given in \cite{Moller:2004qq} are not correct. 
We will provide the corrected expressions.  
Most importantly, in this comment we want to add the second order expansion of 
the hybrid SW map for the Higgs field to 
the work \cite{Alboteanu:2007bp}. This is necessary if one wants to consider 
the Noncommutative Standard Model (NCSM)
\cite{Calmet:2001na,Aschieri:2002mc,Melic:2005fm} including the Yukawa couplings. 
We have computed a special solution and not the most general one, because of 
the complexness. 
The hybrid SW map of the Higgs field was also not discussed in \cite{Moller:2004qq}. 
There, the author considers only $\theta$-expanded Yang-Mills theory coupled 
to matter. Also, other NCSM issues were not addressed in \cite{Moller:2004qq}.
Therefore, in this brief report we present the 2nd order SW $\theta$ expansion
for gauge parameter, vector potential, fermion and Higgs fields, all ingredients 
necessary for complete construction of the NCSM at the 2nd order in $\theta$.

The SW map of the noncommutative gauge parameter $\widehat \Lambda_\alpha$ 
is a solution of the relation
\begin{equation}
i \delta_\alpha \widehat \Lambda_\beta - i\delta_\beta \widehat \Lambda_\alpha + 
[\widehat \Lambda_\alpha\stackrel{*}{,} \widehat \Lambda_\beta] 
= i \widehat \Lambda_{\alpha\times \beta}\,.
\label{sw3}
\end{equation}
This relation has to be solved order by order. 
Therefore, we expand the NC gauge parameter 
$\widehat \Lambda_\alpha$ in $\theta$ and in terms of the commutative gauge parameter $\alpha$,
\begin{eqnarray}
\widehat \Lambda_\alpha&=&\alpha + 
\Lambda_\alpha^{\theta}[V] + \Lambda_\alpha^{\theta^2}[V]+ \mathcal O(\theta^3).
\label{swe1}
\end{eqnarray}
Noncommutative fields and gauge parameters are denoted by a hat throughout the paper.
To first order in $\theta$, the equivalence condition (\ref{sw3}) reads
\begin{eqnarray}
i(\delta_\alpha \Lambda_\beta^\theta - \delta_\beta \Lambda_\beta^ \theta ) &+&
[\alpha, \Lambda_\beta^\theta] + [\Lambda_\alpha^\theta,\beta] 
- i \Lambda^\theta_{\alpha\times \beta} 
\label{sw4}\\
&=&-\frac{i}{2} \theta^{\mu\nu}\{ \partial_\mu \alpha, \partial_\nu \beta\},
\nonumber
\end{eqnarray}
and to second order
\begin{eqnarray}
\Delta \Lambda^2 &=& \frac18 \theta^{\mu\nu}\theta^{\kappa\lambda} 
 [ \partial_\mu\partial_\kappa\alpha, \partial_\nu \partial_\lambda \beta ] 
 - [ \Lambda_\alpha^\theta, \Lambda_\beta^\theta] 
 \nonumber\\
 &-& \frac{i}2 \theta^{\mu\nu}
 \Big(
 \{\partial_\mu \Lambda^\theta_\alpha, \partial_\nu \beta\} - \{\partial_\nu \alpha, 
 \partial_\mu \Lambda^\theta_\beta\}
\Big)\,,
\label{sw5}
\end{eqnarray}
with $\alpha\times \beta = -i[\alpha,\beta]$.
This is an inhomogeneous equation. The homogeneous part to order $k$ is given by
\begin{equation}
\Delta \Lambda^k := i(\delta_\alpha \Lambda_\beta^{\theta^k} 
- \delta_\beta \Lambda_\alpha^{\theta^k})
+ [\alpha,\Lambda_\beta^{\theta^k}] + [\Lambda_\alpha^{\theta^k}, \beta] 
- i\Lambda_{\alpha\times \beta}^{\theta^k} =0.
\label{sw6}
\end{equation}
The most general solution to 1st order, is given by 
\begin{equation}
\Lambda_\alpha^\theta = \frac12 \theta^{\mu\nu} \{ V_\nu, \partial_\mu \alpha \}_c,
\label{sw7}
\end{equation}
where $\{A,B\}_c \equiv c A\cdot B + (1-c) B\cdot A$. 
The requirement of hermiticity fixes the free parameter $c$ to $c=1/2$.
In this case, the general solution (\ref{sw7}) becomes
\begin{eqnarray}
\Lambda_\alpha^{\theta}[V] =
\frac{1}{4} \theta^{\mu \nu}
           \{ V_{\nu}, \partial_{\mu}\alpha\}\,.
	   \label{sw8}
\end{eqnarray}
A special solution for the 2nd order equation reads:
\begin{eqnarray}
\Lambda_\alpha^{\theta^2}[V] &=&
\frac{1}{32}\theta^{\mu\nu}
\theta^{\kappa\lambda}
\Big(\{V_{\mu},\{\partial_{\nu}V_{\kappa},\partial_{\lambda}\alpha\}\}
\nonumber\\
&+&
\{V_{\mu},
	\{V_{\kappa},\partial_{\nu}\partial_{\lambda}\alpha\}\}+         
\{\{V_{\mu},\partial_{\nu}V_{\kappa}\},\partial_{\lambda}\alpha\}\}
\nonumber\\
&-&
\{\{F_{\mu\kappa},V_{\nu}\},\partial_{\lambda}\alpha\}
	-2i[
\partial_{\mu}V_{\kappa},\partial_{\nu}\partial_{\lambda}\alpha] \Big),
\label{sw9}
\end{eqnarray}
with the field strength $F_{\alpha\beta}=\partial_\alpha V_\beta 
- \partial_\beta V_\alpha - i[V_\alpha,V_\beta]$.


The noncommutative gauge field $\widehat V_\mu$ transforms as
\begin{equation}
\delta_\alpha \widehat V_\mu = \partial_\mu \widehat \Lambda_\alpha 
- i[ \widehat V_\mu \stackrel{\star}{,} \widehat \Lambda_\alpha]\,.
\label{sw10}
\end{equation}
The enveloping algebra valued gauge potential is therefore determined by
the following consistency relations in first and second order in $\theta$
\begin{eqnarray}
\label{equivalence1}
\Delta_\alpha V^\theta_\sigma & = & \partial_\sigma  \Lambda_\alpha^\theta - 
	i[V_\sigma, \Lambda^\theta_\alpha] 
	+ \frac12 \theta^{\mu\nu} \{\partial_\mu V_\sigma, \partial_\nu \alpha\},
\\
\label{sw11}
\Delta_\alpha V^{\theta^2}_\sigma & = & \partial_\sigma  \Lambda_\alpha^{\theta^2} - 
	i[V_\sigma, \Lambda^{\theta^2}_\alpha] -i[ V^\theta_\sigma, \Lambda_\alpha^\theta]
	\nonumber\\
	&+& \frac12 \theta^{\mu\nu} \{\partial_\mu V_\sigma^\theta, \partial_\nu \alpha\}
+ \frac12 \theta^{\mu\nu} \{\partial_\mu V_\sigma, \partial_\nu \Lambda_\alpha^\theta \}
\nonumber\\
	&+& \frac{i}8\theta^{\mu\nu } \theta^{\kappa\lambda} 
	[ \partial_\mu\partial_\kappa V_\sigma, \partial_\nu\partial_\lambda \alpha ],
\end{eqnarray} 
where we have again used $\Delta_\alpha V^{\theta^k}_\sigma 
= \delta_\alpha V^{\theta^k}_\sigma - i[\alpha,V^{\theta^k}_\sigma]$
and the $\theta$-expansion of the NC gauge field $\widehat V_\mu$
\begin{equation}
\widehat V_\mu [V] = V_\mu+ V^{\theta}_\mu[V]+ V^{\theta^2}_\mu[V]+\mathcal O(\theta^3)\,.
\label{swe2}
\end{equation}
The general solution in first order in $\theta$ is
\begin{equation}
V^\theta_\mu = \frac12 \theta^{\alpha\beta} \{ V_\beta, \partial_\alpha V_\mu\}_c
		+ \frac12 \theta^{\alpha\beta}\{ V_\beta, F_{\alpha\mu}\}_c \,.
\label{sw12}
\end{equation}
Choosing a hermitian gauge parameter $\widehat\Lambda_\alpha$, we obtain
\begin{eqnarray}
 V_\mu^{\theta}[V]  =  
    \frac{1}{4} \theta^{\alpha \beta}
           \{ \partial_{\alpha}V_{\mu} + F_{\alpha \mu},V_{\beta}\}\,
\label{sw13}
\end{eqnarray}
and
\begin{eqnarray}
V_\mu^{\theta^2}[V] & = & 
	\frac1{64} \, \theta^{\alpha\beta} \theta^{\gamma\delta} 
	\Bigg( 4 [ \partial_{\delta} V_{\beta}, [ V_\mu, \partial_{\alpha} V_{\gamma} ]]
\nonumber\\	
	&+&	
	8 \{ V_{\alpha}, \{ F_{\mu\gamma}, F_{\beta\delta} \} \}
        +8 \{ V_{\alpha}, \{ \partial_{\beta} F_{\mu\gamma}, V_{\delta} \} \} 
\nonumber\\
       &+&2i \{ V_{\alpha}, \{ \partial_\mu V_{\beta}, V_{\gamma} V_{\delta} \} \}
        -2 \{ V_{\alpha}, \{ \partial_{\beta}\partial_\mu V_{\gamma}, V_{\delta} \} \}
\nonumber\\
	&-& \{ V_\mu,  \{ F_{\alpha\gamma}, F_{\beta\delta} \} \} 
	+ 8 \{ \partial_{\alpha} V_\mu, \{ \partial_{\gamma} V_{\beta}, V_{\delta} \} \}
\nonumber\\	
	&+& 2 \{ V_\mu, \{ V_{\alpha} V_{\beta}, V_{\gamma} V_{\delta} \} \}
	+ 2 \{ \partial_{\alpha} V_{\gamma}, \{ V_{\beta}, \partial_\mu V_{\delta} \} \}
\nonumber\\
	&-& 2 \{ \partial_{\mu} V_{\alpha}, \{ F_{\beta\gamma}, V_{\delta} \} \}
	- 2 \{ V_{\alpha} V_{\beta}, \{ V_\mu, V_{\gamma} V_{\delta} \} \}
\nonumber\\	
	&-& 4 \{ V_{\alpha} V_{\gamma}, \{ V_\mu, V_{\beta} V_{\delta} \} \}
	+ 8 \{ V_{\alpha} V_\mu V_{\gamma}, V_{\beta} V_{\delta} \}
\nonumber\\
	&+& 8i [ \partial_{\alpha} \partial_{\gamma} V_\mu, \partial_{\beta} V_{\delta} ]	
	- 2i [ \partial_\mu F_{\alpha\gamma}, F_{\beta\delta} ]
\nonumber\\
	&-& 4i [ \partial_{\alpha} \partial_\mu V_{\gamma}, \partial_{\delta} V_{\beta} ]
	-4  V_{\alpha} (\partial_{\beta} V_{\gamma}) \partial_\mu V_{\delta} 
\nonumber\\
	&-& 4 (\partial_\mu V_{\alpha}) (\partial_{\gamma} V_{\beta}) V_{\delta}
	+ 2i V_{\alpha} V_{\gamma} (\partial_{\beta} V_{\delta}) V_\mu
\nonumber\\	
	&-& 4i V_{\alpha} V_{\gamma} (\partial_\mu V_{\beta}) V_{\delta}
	- 2i V_{\alpha} (\partial_{\beta} V_{\gamma}) V_{\delta} V_{\mu}
\nonumber\\	
	&-& 2i V_{\alpha} (\partial_{\gamma} V_{\beta}) V_{\delta} V_{\mu}
	+ 4i V_{\alpha} (\partial_{\mu} V_{\gamma}) V_{\beta} V_{\delta}
\nonumber\\	
	&-& 2i V_{\mu} V_{\alpha} V_{\gamma} (\partial_{\beta} V_{\delta})
	+ 2i V_{\mu} V_{\alpha} (\partial_{\beta} V_{\gamma}) V_{\delta} 
\nonumber\\
	&+& 2i V_{\mu} V_{\alpha} (\partial_{\gamma} V_{\beta}) V_{\delta}
	- 2i V_{\mu} (\partial_{\alpha} V_{\gamma}) V_{\delta} V_{\beta}
\nonumber\\	
	&+& 2i (\partial_{\alpha} V_{\gamma}) V_{\delta} V_{\beta} V_{\mu}
	+4 F_{\alpha\gamma} V_\mu F_{\beta\delta}\Bigg)\, .
\label{sw14}
\end{eqnarray}
Concerning Eq.~(\ref{sw14}), the second order of the SW map, 
we disagree with ref. \cite{Moller:2004qq}.
The solution given there, does not satisfy the gauge equivalence relation (\ref{sw11}).


Next, to find the Seiberg-Witten map for fermion fields we have to use
the SW map for the NC gauge parameter (\ref{sw4}) to (\ref{sw9}), 
obtained as solution to the relation (\ref{sw3}) via $\theta$-expansion (\ref{swe1}),
together with the noncommutative gauge transformation for the NC fermion fields $\widehat \psi$
\begin{eqnarray}
\delta_\alpha \widehat \psi = i \widehat \Lambda_\alpha  \star \widehat \psi\,.
\label{sw15}
\end{eqnarray}
The $\theta$-expansion of the NC fermion fields $\widehat \psi$ reads
\begin{equation}
\widehat \psi[\psi,V] = \psi+ \psi^{\theta}[\psi,V] +\psi^{\theta^2}[\psi,V]
+ \mathcal O(\theta^3)\,.
\label{swe3}
\end{equation}
The above leads to the 1st order in $\theta$ consistency relation,
\begin{equation}
\Delta_\alpha \psi^\theta = i\Lambda^\theta \psi - \frac12 \theta^{\mu\nu} \partial_\mu
\alpha \partial_\nu \psi,
\label{sw16}
\end{equation}
 where $\Delta_\alpha \psi^{\theta^k}\equiv \delta_\alpha \psi^{\theta^k} 
- i\alpha\psi^{\theta^k}$. 
To 2nd order, we obtain
\begin{eqnarray}
\Delta_\alpha \psi^{\theta^2} &= &i\Lambda_\alpha^{\theta^2} \psi + i\Lambda^\theta \psi^\theta
- \frac12 \theta^{\mu\nu}\partial_\mu \Lambda^\theta_\alpha \partial_\nu\psi
\label{sw17}\\
&-& \frac12 \theta^{\mu\nu}\partial_\mu \alpha \partial_\nu\psi^\theta
-\frac{i}8 \theta^{\mu\nu} \theta^{\kappa\lambda}\partial_\mu\partial_\kappa\alpha
\partial_\nu\partial_\lambda \psi\,.
\nonumber
\end{eqnarray}
The general solution to first order in $\theta$ is given by
\begin{equation}
\psi^\theta = \frac12 \theta^{\mu\nu} V_\nu \partial_\mu \psi + \frac{1-c}2 \theta^{\mu\nu}
 \partial_\mu V_\nu \psi+ d\, \theta^{\mu\nu}F_{\mu\nu}.
 \label{sw18}
\end{equation}
The hermicity requirement $c=1/2$, and choice $d=-1/8$, leaves us with 
\begin{eqnarray}
\psi^{\theta}[\psi,V]  =  - \, \frac{1}{2}\,\theta^{\alpha \beta} \,
\,\Big( V_{\alpha}  \,\partial_{\beta}
          - \frac{i}{4}  \,
              [V_{\alpha}, V_{\beta}] \Big)  \,\psi \, .
\label{sw19}
\end{eqnarray}
A solution to the second order in $\theta$ consistency relation is given by 
\begin{eqnarray}
\psi^{\theta^2}[\psi,V]&=& \frac{1}{32} \,\theta^{\mu\nu}\theta^{\kappa\lambda} 
\Big(-4i (\partial_\kappa V_\mu) \partial_\nu\partial_\lambda  
+4V_\kappa
V_\mu\partial_\nu\partial_\lambda 
\nonumber\\
&-& 4 ( \partial_\kappa V_\mu ) V_\nu \partial_\lambda +4F_{\kappa \mu} V_\nu \partial_\lambda
\nonumber\\
&-& 4 V_\nu ( \partial_\kappa V_\mu )
 \partial_\lambda +8 V_\nu F_{\kappa \mu} \partial_\lambda 
-8i V_\mu V_\kappa V_\nu \partial_\lambda 
\nonumber\\
&+&4i V_\mu V_\nu
V_\kappa\partial_\lambda  -2 ( \partial_\kappa
V_\mu ) \partial_\lambda V_\nu + V_\kappa V_\lambda V_\mu V_\nu 
\nonumber\\
&+&2i ( \partial_\kappa
V_\mu ) V_\lambda V_\nu-2i V_\nu V_\lambda \partial_\kappa V_\mu 
-2 V_\kappa V_\mu V_\nu V_\lambda
\nonumber\\
&-&i[[(\partial_\kappa V_\mu), V_\nu], V_\lambda] - 4i V_\nu F_{\kappa \mu}V_\lambda 
\Big)\,\psi.
\label{sw20}
\end{eqnarray}


Finally we consider noncommutative Higgs field $\widehat{\Phi}$, 
which is related to the commutative ones by 
the hybrid Seiberg-Witten ($\theta$-expansion) map
\begin{equation}
\widehat{\Phi}[\Phi,V,V']
=\Phi+\Phi^{\theta}[V,V']+\Phi^{\theta^2}[V,V']
+ \mathcal O(\theta^3)\,. 
\label{swe4}
\end{equation}
Equation (\ref{swe4}) generalizes the SW maps of both gauge bosons and fermions.
$\widehat\Phi$ is a functional of a commutative Higgs field $\Phi$ and two gauge fields $V,V'$.
It transforms covariantly under the following gauge transformations:
\begin{equation}
\label{sw21}
\delta \widehat\Phi[\Phi,V,V'] = i\widehat \Lambda * \widehat \Phi - i\widehat
\Phi*\widehat \Lambda'\, ,
\end{equation}
where $\widehat\Lambda$ and $\widehat\Lambda'$ are the corresponding NC gauge
parameters. Hermitian conjugation yields
$\widehat\Phi[\Phi,V,V']^\dagger = \widehat\Phi[\Phi^\dagger,V',V]$.
The noncommutative covariant derivative for the NC Higgs field
$\widehat{\Phi}$ is given by
\begin{equation}
\widehat{D}_{\mu} \widehat{\Phi} =
\partial_{\mu} \widehat{\Phi}
-i \, (\widehat{V}_{\mu} * \widehat{\Phi}
- \, \widehat{\Phi} * \widehat{V}'_{\mu})
\, .
\label{sw22}
\end{equation}
As explained in \cite{Calmet:2001na},
the precise representations of the gauge fields
$V$ and $V'$ in the Yukawa couplings
are inherited from the fermions
on the left ($\bar{\psi}$) and on the right side ($\psi$)
of the Higgs, respectively.
The hybrid SW map for the Higgs field up to second order is
of course only unique up to a solution of the homogeneous equation. 
The most general solution up to 1st order in $\theta$ reads
\begin{eqnarray}
\Phi^{\theta}[\Phi,V,V'] &=& 
\frac{1}{2} \, \theta^{\alpha \beta}
\Big[
     V_{\beta}
  \left( \partial_{\alpha} \Phi
       - \frac{i}{2} (a V_{\alpha} \Phi - \Phi V'_{\alpha}) \right)
 \nonumber\\
&+& 
  \left( \partial_{\alpha} \Phi
       - \frac{i}{2} (V_{\alpha} \Phi - b \Phi V'_{\alpha}) \right)
       V'_{\beta}
       \label{sw23}\\
&+& 
\frac14(1-a) (\partial_\alpha V_\beta)\, \Phi 
+ \frac14 (1-b) \Phi \, (\partial_\alpha V'_\beta)
\Big].
\nonumber
\end{eqnarray}
Conventionally, we choose $a=b=1$ and obtain to the first order in $\theta$
\begin{eqnarray}
\Phi^{\theta}[\Phi,V,V'] &=& 
\frac{1}{2} \, \theta^{\alpha \beta}\,
\Big[
     V_{\beta}
  \left( \partial_{\alpha} \Phi
       - \frac{i}{2} (V_{\alpha} \Phi - \Phi V'_{\alpha}) \right)
       \nonumber\\
 &+& 
  \left( \partial_{\alpha} \Phi
       - \frac{i}{2} (V_{\alpha} \Phi - \Phi V'_{\alpha}) \right)
       V'_{\beta}\Big]\,,
\label{sw24}
\end{eqnarray}
while for the 2nd order we have found the following lengthy expression:
\begin{eqnarray}
\Phi^{\theta^2} & \equiv & \Phi^{\theta^2}[\Phi,V,V'] =  
 -\frac{i}{32}\,\theta^{\alpha\beta }  \theta^{\gamma \delta }
 \nonumber\\
&\times & 
 \Bigg(
 V_\alpha  
 \Bigg\{
 V_{\beta }  \big(V_{ \gamma}  \big(
 4 \partial_\delta   \Phi -3   i    V_\delta   {\Phi }+4   i   
{\Phi }   V'_\delta \big)
\nonumber\\
&+&
\big(-4  \partial_\gamma  \Phi -2 
 i    {\Phi }   V'_{\gamma} \big)   V'_\delta \big) 
 +V_{\gamma}  \big[4   i  \partial_\beta \partial_\delta \Phi 
\nonumber\\
&+&
  V_{\beta }  \big(-4  \partial_\delta  \Phi + 2   i    \big(V_\delta  \Phi - 
 2  \Phi    V'_\delta \big)\big)
\nonumber\\ 
&+&
V_\delta  \big(4  \partial_\beta \Phi + 4   i    \Phi    V'_\beta \big)
\nonumber\\
&+&
3 ( \partial_\beta V_\delta ) \Phi - 4 ( \partial_\beta \Phi ) V'_\delta 
-4  \partial_\delta \Phi   V'_\beta 
\nonumber\\ 
&+& 
\Phi   \big( 4\partial_\delta  V'_\beta - 8  \partial_\beta  V'_\delta  
+4 i \big( V'_\beta  V'_\delta -2 V'_\delta  V'_\beta \big)\big)\big]
\nonumber\\
&+& \partial_\beta V_\gamma
 \big(8   i    \partial_\delta \Phi + 5   V_\delta  
\Phi - 8  \Phi    V'_\delta \big) 
\nonumber\\
&+& 
\partial_\gamma V_\beta  
\big(-4   i   \partial_\delta  \Phi -3  V_\delta   \Phi \big)
\nonumber\\
&+&
4 \partial_\gamma  \Phi  \big(  - i \partial_\beta V'_\delta + i  \partial_\delta V'_\beta 
+ V'_\beta
  V'_\delta + V'_\delta   V'_\beta \big)
\nonumber\\  
&+&
\big(-8   i 
  \partial_\beta \partial_\gamma  \Phi + 4 ( \partial_\beta
\Phi ) V'_\gamma \big) V'_\delta
\nonumber\\
&+& 
4\Phi \big[ V'_\gamma  \big(  \partial_\beta V'_\delta -  \partial_\delta  V'_\beta -  i  V'_\beta
V'_\delta +  i   V'_\delta  V'_\beta \big)
\nonumber\\
&+&
\big(2 
\partial_\gamma   V'_\beta +   i   V'_\beta 
 V'_\gamma \big) V'_\delta \big]
 \Bigg\}  
\nonumber\\
&+&
\partial_\alpha  V_\gamma  \Big[4  \partial_\beta  \partial_\delta  \Phi 
 +V_{\beta }  \big(-4   i  \partial_\delta \Phi + 4  \Phi    V'_\delta \big)
\nonumber\\
& - &
V_\delta   \big( V_\beta \Phi + 4  \Phi    V'_\beta \big) + 4 i ( \partial_\beta
\Phi ) V'_\delta
\nonumber\\
&-&
2i ( \partial_\delta V_\beta )  
\Phi - 4i ( \partial_\delta \Phi ) V'_\beta  
\nonumber\\
&+& 
\Phi \big( 4   i  \partial_\delta   V'_\beta - 4 
 V'_\beta    V'_\delta + 8   V'_\delta   V'_\beta \big)\Big]
\nonumber\\
&+&
\partial_\alpha \Phi \Big[ V'_\gamma   \big(-4   i 
 \partial_\delta  V'_\beta + 4   V'_\beta    V'_\delta -4   V'_\delta  V'_\beta \big)
\nonumber\\
&+&
\big( 8 i \partial_\gamma V'_\beta - 4 i \partial_\beta V'_\gamma 
- 4 V'_\beta V'_\gamma \big) V'_\delta \Big]
\nonumber\\
&+&
\partial_\alpha  \partial_\gamma  \Phi   \big(
4i V'_\delta V'_\beta - 4 \partial_\delta  V'_\beta \big)  
\nonumber\\
&+&
\Phi  \Bigg\{ V'_\alpha   \big[ V'_\gamma  
\big( \partial_\beta V'_\delta +2   i     V'_\beta   V'_\delta \big) 
\nonumber\\
&+&
\big(3  \partial_\beta V'_\gamma -5\partial_\gamma V'_\beta -3  i V'_\beta V'_\gamma \big)   
V'_\delta \big] 
\nonumber\\
&+& 
\partial_\alpha V'_\gamma   \big(-2   i \partial_\delta  V'_\beta - 3  V'_\delta   
V'_\beta \big)\Bigg\}
\Bigg)\,.
\label{sw25}
\end{eqnarray}
The above expression may be written in a more convenient way as
\begin{eqnarray}
&&\Phi^{\theta^2} [\Phi, V,V']  =  
\Phi^{\theta^2}[V] + \Phi^{\theta^2}_r[V']
-\frac{i}{8}   \theta^{\mu \nu }  \theta^{\kappa \lambda } 
\nonumber\\
&&\times
\Bigg(
 i V_\kappa  V_{\lambda } V_{ \mu}   {\Phi }   V'_\nu 
 -  V_\kappa  V_{\lambda } (\partial_\mu  \Phi) V'_\nu
 - i V_\kappa  V_{ \mu} V_{\lambda } \Phi  V'_\nu 
\nonumber\\
&&+   
i  V_\kappa  V_{ \mu} V_\nu  \Phi    V'_\lambda 
-  V_\kappa  V_{ \mu} (\partial_\lambda \Phi ) V'_\nu  
-  V_\kappa  V_{ \mu} ( \partial_\nu \Phi )  V'_\lambda   
\nonumber\\
&&- 
2 V_\kappa  V_{ \mu} \Phi (\partial_\lambda  V'_\nu ) 
+  V_\kappa  V_{ \mu} \Phi (\partial_\nu  V'_\lambda) 
+  i V_\kappa  V_{ \mu} \Phi V'_\lambda    V'_\nu 
\nonumber\\
&&-
 2 i V_\kappa  V_{ \mu} \Phi V'_\nu    V'_\lambda
- 2 V_\kappa  (\partial_\lambda V_\mu)  \Phi    V'_\nu 
\nonumber\\
&&-
 4 i V_\kappa  (\partial_\mu  \Phi ) \partial_\lambda V'_\nu
+ i V_\kappa  (\partial_\mu  \Phi ) \partial_\nu V'_\lambda 
\nonumber\\
&&+
 V_\kappa  (\partial_\mu  \Phi ) V'_\lambda V'_\nu 
 +  V_\kappa  (\partial_\mu  \Phi ) V'_\nu    V'_\lambda 
\nonumber\\
&&- 
2 i V_\kappa  ( \partial_\lambda  \partial_\mu  \Phi ) V'_\nu
 +  V_\kappa  ( \partial_\lambda \Phi )  V'_\mu V'_\nu 
 +   V_\kappa  \Phi V'_\mu \partial_\lambda V'_\nu 
 \nonumber\\
&&-  
V_\kappa  \Phi V'_\mu \partial_\nu   V'_\lambda  
-  i V_\kappa  \Phi V'_\mu V'_\lambda V'_\nu
+  i V_\kappa  \Phi V'_\mu V'_\nu  V'_\lambda 
\nonumber\\
&&+
 2 V'_\kappa  \Phi ( \partial_\mu  V'_\lambda ) V'_\nu
 +  i V_\kappa  \Phi V'_\lambda V'_\mu V'_\nu 
 +  (\partial_\kappa  V_\mu) V_{\lambda } \Phi    V'_\nu 
\nonumber\\
&&-
  (\partial_\kappa  V_\mu) V_\nu \Phi  V'_\lambda 
  + i (\partial_\kappa  V_\mu) (\partial_\lambda \Phi )  V'_\nu 
\nonumber\\
&&-
 i  (\partial_\kappa  V_\mu) ( \partial_\nu \Phi ) V'_\lambda 
+ i (\partial_\kappa  V_\mu) \Phi \, \partial_\nu V'_\lambda
\nonumber\\
&&-
 (\partial_\kappa  V_\mu) \Phi V'_\lambda    V'_\nu 
 + 2 (\partial_\kappa  V_\mu) \Phi V'_\nu V'_\lambda 
\Bigg)
\nonumber\\
&&-
\frac1{32} \theta^{\mu \nu }  \theta^{\kappa \lambda }
 2 V_\kappa  V_{\lambda } {\Phi } V'_{\mu} V'_\nu \,,
 \label{sw26}
\end{eqnarray}
where $\Phi^{\theta^2}[V]$ and $\Phi^{\theta^2}_r[V']$ 
denote the second order expansion for fermion fields (\ref{sw20}),
\begin{eqnarray}
\Phi^{\theta^2}[V]&=& \psi^{\theta^2}[\psi,V](\psi \to \Phi)\,, 
\label{sw27}
\end{eqnarray}
in the latter case the gauge fields are supposed to act from the right,
\begin{eqnarray}
\Phi_r^{\theta^2}[V]&=&\,\frac{1}{32} \,\theta^{\mu\nu}\theta^{\kappa\lambda} 
\nonumber\\
&\times& \Phi
\Big( - 4i (\partial_\kappa
V_\mu ) \partial_\nu\partial_\lambda  
+4V_\kappa
V_\mu\partial_\nu\partial_\lambda 
\nonumber\\
&-& 4 (\partial_\kappa
V_\mu ) V_\nu \partial_\lambda +4F_{\kappa \mu} V_\nu \partial_\lambda
\nonumber\\
&-& 4 V_\nu ( \partial_\kappa
V_\mu ) \partial_\lambda +8 V_\nu F_{\kappa \mu} \partial_\lambda 
-8i V_\mu V_\kappa V_\nu \partial_\lambda 
\nonumber\\
&+&4i V_\mu V_\nu
V_\kappa\partial_\lambda  -2\partial_\kappa
V_\mu\partial_\lambda V_\nu + V_\kappa V_\lambda V_\mu V_\nu 
\nonumber\\
&+& 2i ( \partial_\kappa
V_\mu ) V_\lambda V_\nu-2i V_\nu V_\lambda \partial_\kappa V_\mu 
-2 V_\kappa V_\mu V_\nu V_\lambda
\nonumber\\
&-&i[[(\partial_\kappa V_\mu), V_\nu], V_\lambda] - 4i V_\nu F_{\kappa \mu}V_\lambda 
\Big)\,.
\label{sw28}
\end{eqnarray}

The solutions representing SW maps up 
to second order in $\theta$ for fermion fields and Higgs 
fields, in the case of $V'=0$, are the same, 
i.e. the first order in $\theta$ Eqs. (\ref{sw19}) and (\ref{sw24}) becomes identical. 
The same holds for the second order Eqs. (\ref{sw20}) and (\ref{sw25}) or (\ref{sw26}).

Higher SW expansions of the NC gauge, fermion and Higgs fields
up to second order in the noncommutative parameter $\theta$, 
are important due to the further extension of previously published results. 
Specifically those are applications of the enveloping algebra based, 
$\theta$-expanded, approach to
higher gauge groups \cite{Aschieri:2002mc} and to particular NCSM gauge sector
representations \cite{Goran}. Proof that SW noncommutative gauge theories
are anomaly free and the properties of the gauge anomaly for general SW mapping
as well as of the U(1)$_{\rm A}$ anomaly in noncommutative 
SU(N) theories \cite{Martin:2002nr}
are certainly very important.
This comment is important regarding further investigations 
of renormalizability properties 
of the $\theta$-expanded NC field theories in general \cite{Grosse:2005da}. 
Certainly, recent results \cite{Maja,Martin:2006gw}, 
showing that gauge theories, in the $\theta$-expanded 
enveloping algebra based approach, are one-loop renormalizable
at first order in $\theta$, are very encouraging. 
They give us hope that it would be possible 
to investigate higher-loop renormalizability up to   
the second order in the noncommutative parameter $\theta$. 
Clearly, one may expect that the renormalizability principle should certainly
help to minimize, or even cancel most of ambiguities of SW maps discussed in 
\cite{Alboteanu:2007bp,Moller:2004qq} and in this comment.
Finally, it is necessary to comment that, 
due to the one-loop renormalizability 
\cite{Grosse:2005da,Maja,Martin:2006gw}, the associated
high energy particle physics phenomenology 
\cite{Goran,Josip,Ohl:2004tn} becomes more robust \cite{Buric:2006nr}. 
\\


We want to thank Fabian Bachmaier for contributing to this work 
and to H. Grosse and J. Wess for many fruitful discussions. 
M.W. also wants to acknowledge the support from ``Fonds zur F\"orderung der
wissenschaftlichen Forschung" (Austrian Science Fund), project
P18657-N16. The work of J.T. is supported by the project 098-0982930-2900 of
the Croatian Ministry of Science, Education and Sport and in part by
the ESF, received in the framework of the Research
Networking Programme on 'Quantum Geometry and Quantum Gravity' in the form of
a short visit grant.


\begin{thebibliography}{99}

\bibitem{Alboteanu:2007bp}
A.~Alboteanu, T.~Ohl and R.~R\"uckl, 
0707.3595[hep-th].

\bibitem{Moller:2004qq}
L.~M{\"o}ller, 
JHEP {\bf 10} (2004) 063.

\bibitem{Seiberg:1999vs}
N. Seiberg and E. Witten,
JHEP {\bf 09} (1999) 032.

\bibitem{Madore:2000en}
J. Madore, S. Schraml, P.~Schupp and J.~Wess,
Eur. Phys. J. C{\bf 16} (2000) 161.

\bibitem{Jurco:2000ja}
B. Jur\v{c}o, S. Schraml, P.~Schupp and J.~Wess,
Eur. Phys. J. C{\bf 17} (2000) 521.
B. Jur\v{c}o, L. M\"oller, S.~Schraml, P.~Schupp and J.~Wess,
Eur. Phys. J. C{\bf 21} (2001) 383.

\bibitem{Calmet:2001na}
X.~Calmet, B.~Jur\v{c}o, P.~Schupp, J.~Wess and M.~Wohlgenannt,
Eur.~Phys.~J. C{\bf 23} (2002) 363.
%
\bibitem{Aschieri:2002mc}
P. Aschieri, B.~Jur\v{c}o, P.~Schupp and J.~Wess,
Nucl. Phys. B{\bf 651} (2003) 45.
%
\bibitem{Melic:2005fm}
B.~Melic, K.~Passek-Kumericki, J.~Trampetic, P.~Schupp, and M.~Wohlgenannt,
  {\em Eur. Phys. J.} {\bf C42} (2005) 483--497;
  {\em Eur. Phys. J.} {\bf C42} (2005) 499--504.
%
\bibitem{Goran}
W.~Behr, N.G. Deshpande, G. ~Duplan\v{c}i\'{c}, P.~Schupp,
J.~Trampeti\'{c} and J.~Wess,
Eur. Phys. J. C{\bf 29} (2003) 441;
%
G.~Duplan\v{c}i\'{c}, P.~Schupp and J.~Trampeti\'{c},
Eur.~Phys. J. C{\bf 32} (2003) 141.
%
\bibitem{Martin:2002nr}
  C.~P.~Martin,
  Nucl.\ Phys.\  B {\bf 652}, 72 (2003);
  F.~Brandt, C.~P.~Martin and F.~R.~Ruiz,
  JHEP {\bf 0307}, 068 (2003);
  C.~P.~Martin and C.~Tamarit,
  Phys.\ Rev.\  D {\bf 72}, 085008 (2005).
\bibitem{Grosse:2005da}
  %
  R.~Wulkenhaar,
  JHEP {\bf 0203} (2002) 024;
  %
  J.~M.~Grimstrup and R.~Wulkenhaar,
  Eur.\ Phys.\ J.\ C {\bf 26} (2002) 139;
  %
A.~Bichl, J. Grimstrup, H. Grosse, L. Popp, M. Schweda and R.
Wulkenhaar, 
JHEP {\bf 06} (2001) 013;
%
  H.~Grosse and R.~Wulkenhaar,
  Lett.\ Math.\ Phys.\  {\bf 71}, 13 (2005);
  J.\ Nonlin.\ Math.\ Phys.\  {\bf 11S1}, 9 (2004);
  Commun.\ Math.\ Phys.\  {\bf 256}, 305 (2005);
  V.~Rivasseau, F.~Vignes-Tourneret and R.~Wulkenhaar,
  Commun.\ Math.\ Phys.\  {\bf 262}, 565 (2006);
  H.~Grosse and M.~Wohlgenannt,
  {\it Eur. Phys. J.} {\bf C52} (2007) 435--450.
%
\bibitem{Maja}
  M.~Buric and V.~Radovanovic,
  JHEP {\bf 0210} (2002) 074;
%
  JHEP {\bf 0402} (2004) 040;
%
  Class.\ Quant.\ Grav.\  {\bf 22} (2005) 525;
  M.~Buric, D.~Latas and V.~Radovanovic,
  JHEP {\bf 0602} (2006) 046;
 %
  %
  M.~Buric, V.~Radovanovic and J.~Trampetic,
  JHEP {\bf 0703} (2007) 030;
  D.~Latas, V.~Radovanovic and J.~Trampetic,
  Phys.\ Rev.\ D {\bf 76} (2007) 085006.
  %
   \bibitem{Martin:2006gw}
  C.~P.~Martin, D.~Sanchez-Ruiz and C.~Tamarit,
  JHEP {\bf 0702} (2007) 065;
  %
  C.~P.~Martin and C.~Tamarit,
  arXiv:0706.4052 [hep-th].
%
  \bibitem{Josip}
J. Trampeti\'c, 
Acta Phys. Polon.
B{\bf 33} (2002) 4317 [hep-ph/0212309];
%
  P.~Schupp, J.~Trampetic, J.~Wess and G.~Raffelt,
  Eur.\ Phys.\ J.\  C {\bf 36} (2004) 405;
  P.~Minkowski, P.~Schupp and J.~Trampetic,
  Eur.\ Phys.\ J.\  C {\bf 37} (2004) 123;
  B.~Melic, K.~Passek-Kumericki and J.~Trampetic,
  Phys.\ Rev.\ D {\bf 72} (2005) 054004,
 ibid 057502.
 %
  \bibitem{Ohl:2004tn}
T.~Ohl and J.~Reuter, 
Phys. Rev. D{\bf 70} (2004);
  A.~Alboteanu, T.~Ohl and R.~R\"uckl,
  PoS {\bf HEP2005} (2006) 322
  [arXiv:hep-ph/0511188];
  Phys.\ Rev.\ D {\bf 74}, 096004 (2006);
  arXiv:0709.2359 [hep-ph].
  %
\bibitem{Buric:2006nr}
  M.~Buric, D.~Latas, V.~Radovanovic and J.~Trampetic,
  Phys.\ Rev.\ D {\bf 75} (2007) 097701;
  %
 J.~Trampeti\'c,
  arXiv:0704.0559v1 [hep-ph].
  %
%
%
\end{thebibliography}
\end{document}